\documentclass[twocolumn,superscriptaddress,nobibnotes,footinbib,floatfix,aps,prl,citeautoscript,reprint]{revtex4-1}

\usepackage{graphicx}
\usepackage{epsfig}
\usepackage{subfigure}
\usepackage{color}
\usepackage{latexsym}
\usepackage{amsmath,bm,multirow}
\usepackage{amsfonts}
\usepackage{dcolumn}           
\usepackage{natbib}
\usepackage[colorlinks,linkcolor=blue,citecolor=green]{hyperref}
\usepackage{physics}

\usepackage[draft]{changes} 
\usepackage[normalem]{ulem}
\definechangesauthor[name={yi}, color=blue]{yi}

\newcommand{\portland}{Department of Mechanical and Materials Engineering, Portland State University, Portland, OR 97201, USA}

\usepackage[draft]{changes} 
\usepackage[normalem]{ulem}
\usepackage{soul}
\begin{document}
\title{First-Principles Theory of Five- and Six-Phonon Scatterings}

\author{Yi Xia}
\email{yimaverickxia@gmail.com; yxia@pdx.edu}
\affiliation{\portland}

\date{\today}

\begin{abstract}

Higher-order phonon scatterings beyond fourth order remain largely unexplored despite their potential importance in strongly anharmonic materials at elevated temperatures. We develop a theoretical formalism for first-principles calculation of five- and six-phonon scatterings using Green's function techniques based on a diagrammatic formalism, and systematically investigate multi-phonon interactions in Si, MgO, and BaO from room temperature to near melting points. Our calculations reveal dramatically different material-dependent behaviors: while five- and six-phonon processes remain negligible in Si even at high temperatures, they become increasingly important in MgO near its melting point (3100~K) and in BaO at intermediate temperatures (1200~K). Most remarkably, five- and six-phonon scatterings surpass three- and four-phonon scattering intensity in BaO near its melting point (2100~K), reducing lattice thermal conductivity by over 50\%. We demonstrate that the strength of higher-order interactions is primarily governed by interatomic force constants, with BaO exhibiting five- and six-phonon scattering rates over one order of magnitude stronger than MgO despite identical crystal structures, due to large scattering phase space arising from softened harmonic interactions. Our work provides theoretical insights into the lattice dynamics and thermal transport in strongly anharmonic materials and at elevated temperatures.

\end{abstract}

\maketitle

Atomic vibrations significantly influence a material's thermal, electronic, and mechanical properties, with profound implications for lattice thermal transport~\cite{mills1995heat,zheng2021advances}, electron-phonon interactions~\cite{ziman1960electrons,giustino2017electron}, and the development of next-generation thermal management materials~\cite{Bell1457}. In crystalline solids, these vibrations are elegantly described through phonons—quantized collective excitations that traditionally begin with the harmonic approximation~\cite{wallace1998thermodynamics}, treating atomic displacements as simple oscillators. However, this idealized framework fails to capture the rich physics of real materials, which requires the inclusion of anharmonic effects arising from higher-than-2nd-order atomic interactions~\cite{Cowley1968, wallace1998thermodynamics, srivastava1990physics}.

Recent years have witnessed remarkable progress in anharmonic lattice dynamics and thermal transport simulations from first-principles~\cite{esfarjani2014modeling,LINDSAY2018106, McGaughey2019}, which was initially focused on three-phonon (3ph) scattering processes~\cite{maradudin1962scattering,Cowley1964} that dominate lattice thermal conductivity ($\kappa_{\rm L}$)~\cite{peierls1997kinetic} in several materials~\cite{broido2007intrinsic,esfarjani2011heat,garg2011role,li2012thermal,Lindsay2013,tadano,togo2015distributions,tian2012phonon,Lee:2014aa}. The field has advanced considerably with the explicit calculation of four-phonon (4ph) scattering~\cite{Tianli2016}, revealing their critical importance in both weakly~\cite{Tianli2017, kundu2024electron} and strongly anharmonic materials~\cite{pbte2018,ravichandran2018unified,xie2020first}. Subsequent theoretical developments have demonstrated the equally important role of anharmonicity-induced phonon frequency renormalization~\cite{rczbprx, Tadano2015}, as well as a unified thermal transport framework beyond the Boltzmann transport equation~\cite{Simoncelli2019,isaeva2019modeling,simoncelli2022wigner}. Despite these achievements, our understanding of multi-phonon interactions beyond fourth order remains nascent. Preliminary investigations suggest that fifth- and even higher-order phonon scatterings may play unexpected roles in extreme conditions and exotic materials~\cite{yang2022evidence}, yet their explicit calculation has remained elusive due to the lack of theoretical formalisms and the prohibitive scaling of computational demands.

\begin{figure}[htp]
	\includegraphics[width = 0.95\linewidth]{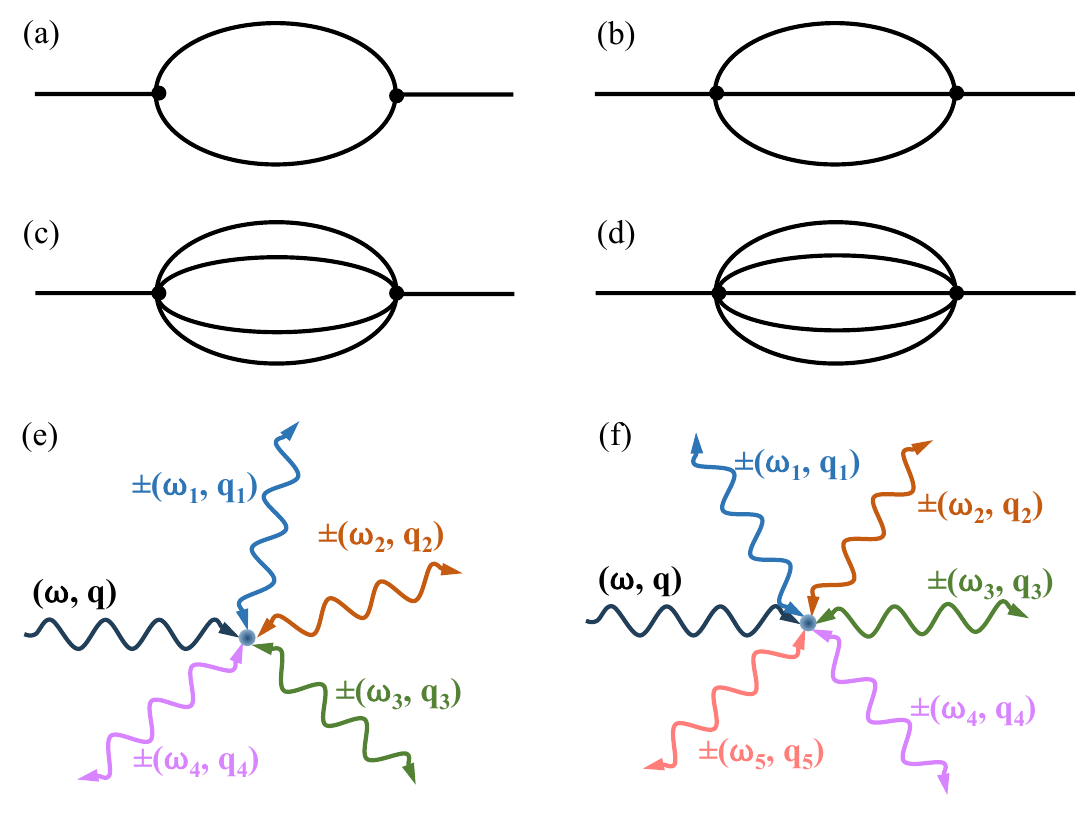}
	\caption{
	Feynman diagrams illustrating (a) three-phonon (3ph), (b) four-phonon (4ph), (c) five-phonon (5ph), and (d) six-phonon (6ph) interactions. Panels (e) and (f) depict schematic representations of 5ph and 6ph scattering processes, respectively, highlighting various combinations of incoming and outgoing phonons.
	}
	\label{fig:diagram}
\end{figure}

\begin{figure*}[htp]
	\includegraphics[width = 1.0\linewidth]{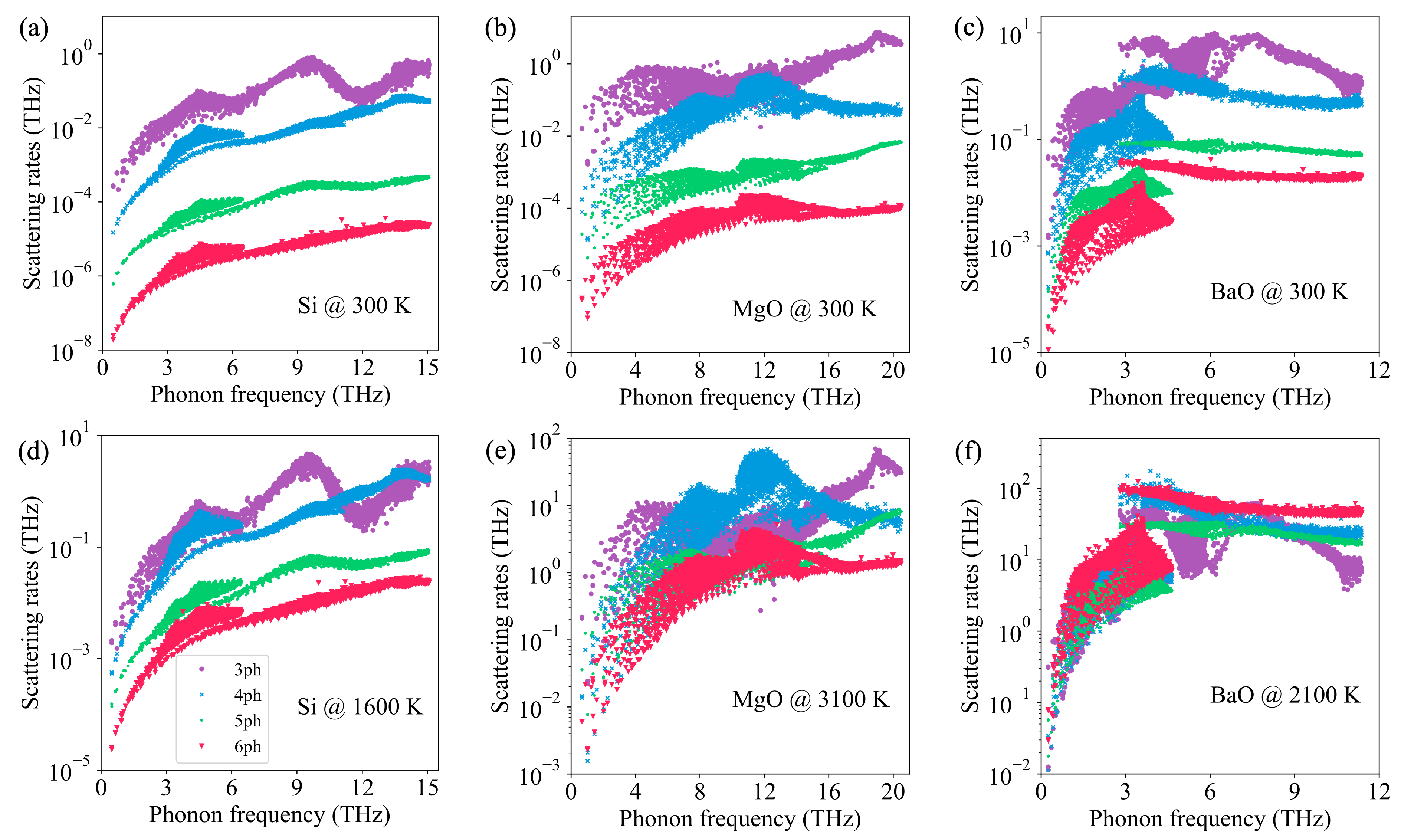}
	\caption{ 
	(a) Calculated phonon mode-resolved scattering rates associated with three-phonon (3ph), four-phonon (4ph), five-phonon (5ph), and six-phonon (6ph) processes in silicon at 300~K. (b) Same as (a), but for MgO. (c) Same as (a), but for BaO. (d)–(f) Corresponding scattering rates at temperatures near the melting points: 1600~K for Si, 3100~K for MgO, and 2100~K for BaO. The two distinct groups of modes in BaO, observed in the scattering rates, are found to correspond to acoustic and optical modes, respectively.
	}
	\label{fig:rates}
\end{figure*}


The first objective of this study is to develop theoretical formalisms for the first-principles calculation of multiphonon scatterings involving up to six-phonon interactions. We note that the quantum field theoretical framework for describing anharmonic phonons was established more than half a century ago through the pioneering work of Cowley~\cite{Cowley1968}, Maradudin and Fein~\cite{maradudin1962scattering}, and Tripathi and Pathak~\cite{tripathi1974self}, amongst others~\cite{della1992equation}. By employing temperature-dependent Green's functions as phonon propagators within the framework of diagrammatic perturbation theory~\cite{abrikosov2012methods}, a systematic evaluation of multiphonon scattering processes can be achieved through the calculation of phonon self-energies, wherein the corresponding Feynman diagrams can be analyzed using established diagrammatic rules~\cite{maradudin1962scattering}

It was found that a special class of phonon Feynman diagrams known as double-vertex diagrams contribute significantly to the phonon self-energy~\cite{della1992equation, procacci1992anharmonic}, as illustrated in Fig.\ref{fig:diagram}(a)-(d). The imaginary parts of these self-energies correspond to multi-phonon scattering rates, with diagrams (a) and (b) representing the well-established 3ph and 4ph scatterings, respectively, while diagrams (c) and (d) represent the previously unexplored five-phonon (5ph) and six-phonon (6ph) scattering processes. Following the theoretical framework established by Tripathi and Pathak~\cite{tripathi1974self} for deriving 3ph/4ph linewidth ($\Gamma_{\rm 3ph/4ph}$) and employing the diagrammatic evaluation rules of Maradudin and Fein~\cite{maradudin1962scattering}, we derive the formulas for evaluating  5ph/6ph linewidth ($\Gamma_{\rm 5ph/6ph}$) as given in Eq.~(\ref{eq:g5ph}) and Eq.(\ref{eq:g6ph}) (see detailed derivations provided in the Supplemental Material~\cite{SM}), wherein $V$ represents the 
anharmonic coupling tensor, $\omega$ denotes the phonon frequency, $n$ is the phonon population following Bose-Einstein statistics, and ($q, j$) serves as the composite index for phonon wave vector and branch, respectively. The different types of 5ph and 6ph scatterings involving up- and down-conversion processes are graphically illustrated in Fig.~\ref{fig:diagram}(e) and (f), respectively.


\begin{widetext}
\begin{equation}
\label{eq:g5ph}
\begin{split}
\Gamma_{\rm 5ph}(q,j) = \frac{600\pi}{\hbar^2} \sum_{q_1,j_1}\sum_{q_2,j_2}\sum_{q_3,j_3}\sum_{q_4,j_4} 
\left| V(q,j; q_1,j_1; q_2,j_2; q_3,j_3; q_4,j_4) \right|^2 & \\
\{ 
\left[ (n_1+1)(n_2+1)(n_3+1)(n_4+1)-n_1n_2n_3n_4 \right] \left[
\delta(\omega-\omega_1-\omega_2-\omega_3-\omega_4) -\delta(\omega+\omega_1+\omega_2+\omega_3+\omega_4) \right] + \\
4\left[ n_1(n_2+1)(n_3+1)(n_4+1) -(n_1+1)n_2n_3n_4 \right] \left[ \delta(\omega+\omega_1-\omega_2-\omega_3-\omega_4) - \delta(\omega-\omega_1+\omega_2+\omega_3+\omega_4)\right]+ \\
3\left[ n_1n_2(n_3+1)(n_4+1) -(n_1+1)(n_2+1)n_3n_4 \right] \left[ \delta(\omega+\omega_1+\omega_2-\omega_3-\omega_4) - \delta(\omega-\omega_1-\omega_2+\omega_3+\omega_4)\right]
\}
\end{split}
\end{equation}	
\end{widetext}

Having established the theoretical formalism, we proceed to the second objective of this study: systematic first-principles calculations of representative materials to investigate beyond-fourth-order phonon scatterings. We selected face-centered cubic silicon (Si), MgO, and BaO as prototype systems: Si represents a technologically important semiconductor with known 4ph contributions at elevated temperatures~\cite{Tianli2017}, MgO constitutes a key Earth's mantle material where quantitative understanding of thermal transport near melting conditions is critical~\cite{tang2010lattice,kwon2020dominant}, and BaO adopts identical crystal system as MgO but exhibits enhanced anharmonicity~\cite{rczbprx}. However, 5ph and 6ph scatterings in these systems remain unexplored. 

\begin{widetext}
\begin{equation}
\label{eq:g6ph}
\begin{split}
\Gamma_{\rm 6ph}(q,j) = \frac{4320\pi}{\hbar^2} \sum_{q_1,j_1}\sum_{q_2,j_2}\sum_{q_3,j_3}\sum_{q_4,j_4}\sum_{q_5,j_5} 
\left| V(q,j; q_1,j_1; q_2,j_2; q_3,j_3; q_4,j_4; q_5,j_5) \right|^2 & \\
\{ 
\left[ (n_1+1)(n_2+1)(n_3+1)(n_4+1)(n_5+1)-n_1n_2n_3n_4n_5 \right] \\ \left[
\delta(\omega-\omega_1-\omega_2-\omega_3-\omega_4-\omega_5) - \delta(\omega+\omega_1+\omega_2+\omega_3+\omega_4+\omega_5) \right] + \\
5\left[ n_1(n_2+1)(n_3+1)(n_4+1)(n_5+1) -(n_1+1)n_2n_3n_4n_5 \right] \\ \left[ \delta(\omega+\omega_1-\omega_2-\omega_3-\omega_4-\omega_5) - \delta(\omega-\omega_1+\omega_2+\omega_3+\omega_4+\omega_5)\right]+ \\
10\left[ n_1n_2(n_3+1)(n_4+1)(n_5+1) -(n_1+1)(n_2+1)n_3n_4n_5 \right] \\ \left[ \delta(\omega+\omega_1+\omega_2-\omega_3-\omega_4-\omega_5) - \delta(\omega-\omega_1-\omega_2+\omega_3+\omega_4+\omega_5)\right]
\}
\end{split}
\end{equation}
\end{widetext}


To evaluate Eq.~(\ref{eq:g5ph}) and (\ref{eq:g6ph}), we performed first-principles calculations using density functional theory (DFT)~\cite{dft,KohnXC} to obtain interatomic force constants (IFCs) up to the 6th order using the Compressive Sensing Lattice Dynamics (CSLD) approach~\cite{csld,csldlong}. We refer the readers to the Supplemental Material~\cite{SM} for more details of our approach and DFT calculations~\footnote{Our DFT calculations were performed using the Vienna {\it Ab\ Initio\/} Simulation Package (VASP)~\cite{Vasp1, Vasp2, Vasp3, Vasp4}, which employed the projector-augmented wave (PAW)~\cite{paw} method in conjunction with the Perdew-Burke-Ernzerhof version of the generalized gradient approximation (GGA)~\cite{gga} for the exchange-correlation functional~\cite{dft}. We extracted harmonic interatomic force constants (IFCs) using the finite-displacement approach implemented in Phonopy~\cite{Togo20151}. We have implemented Eq.(1) and (2) in the main text within the ShengBTE package~\cite{shengbte,wuli2012}}. Given the computational expense of higher-order phonon scattering processes, we employed stochastic sampling of scattering events~\cite{guo2024sampling} and developed a phonon wave vector subsampling approach (see Supplemental Material~\cite{SM}) to balance computational efficiency with statistical precision. For all three materials, the multi-phonon scattering rates were computed using a uniform phonon wave vector mesh of 32$\times$32$\times$32 sampling points in the first Brillouin zone.

\begin{figure*}[htp]
	\includegraphics[width = 1.0\linewidth]{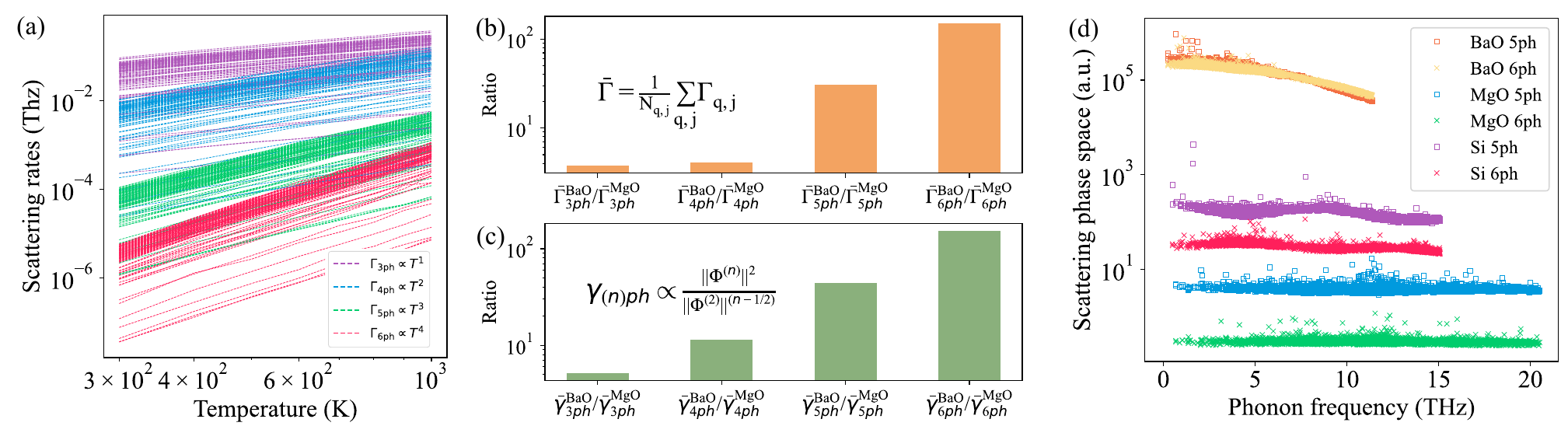}
	\caption{
	(a) Temperature dependencies of three-phonon (3ph), four-phonon (4ph), five-phonon (5ph), and six-phonon (6ph) scattering rates in Si. (b) Mode-averaged 3ph, 4ph, 5ph, 6ph scattering rate ratios of BaO and MgO at 300~K. (c) Interatomic force constants-estimated 3ph, 4ph, 5ph, 6ph scattering rate ratios of BaO and MgO at 300~K. (d) Scattering phase space (as defined in Eqs.~(\ref{eq:g5ph}) and (\ref{eq:g6ph}) with $|V| = 1$) associated with 5ph and 6ph scattering processes in Si, MgO, and BaO at 300~K, respectively.
	}
	\label{fig:ana}
\end{figure*}

We first examine multi-phonon scatterings in Si, MgO, and BaO at room temperature. From Fig.~\ref{fig:rates}(a), we see that Si exhibits significantly weaker higher-than-4th-order phonon scatterings across all phonon frequencies, with 5ph/6ph scatterings more than two orders of magnitude weaker than 3ph/4ph processes. In MgO, although 5ph and 6ph scatterings remain substantially weak, notable overlap between 4ph and 5ph scattering rates is observed in the low-frequency acoustic modes, as shown in Fig.~\ref{fig:rates}(b). For BaO, the strength disparity between 3ph/4ph and 5ph/6ph scattering is further weakened, indicating stronger higher-order anharmonicity (Fig.~\ref{fig:rates}(c)). Overall, 5ph and 6ph scatterings remain relatively weak at room temperature, though their relative strengths compared to 3ph and 4ph scatterings vary significantly among different materials.

Higher-order phonon scatterings are expected to strengthen considerably at elevated temperatures, motivating our investigation of scattering phenomena at near-melting temperatures, as shown in Fig.~\ref{fig:rates}(d)-(f). For Si near its melting temperature, 4ph scattering becomes comparable to 3ph scattering; however, 5ph and 6ph processes remain substantially weaker, though the strength gap between 5ph and 6ph scattering is significantly reduced. In MgO at high temperatures, 5ph and 6ph scattering rates become comparable to each other and begin approaching the magnitudes of 3ph and 4ph scatterings, especially in the acoustic modes. Most notably, in BaO near its melting point, 5ph and 6ph scatterings become extremely strong, even surpassing 3ph and 4ph scatterings in intensity, demonstrating the critical importance of higher-order phonon scatterings in accurately describing phonon dynamics in BaO at elevated temperatures.


By comparing the multi-phonon scattering rates at room and higher temperatures across all three materials, we find that higher-order scattering increases much more rapidly than the lower-order ones. This can be understood from Eq.~[\ref{eq:g5ph}] and [\ref{eq:g6ph}], which reveal that $\Gamma_{\rm 5ph} \propto n^3$ and $\Gamma_{\rm 6ph} \propto n^4$. We illustrate the temperature dependence of 3ph, 4ph, 5ph, and 6ph scatterings in Fig.~\ref{fig:ana}(a) for Si. Consistent with above analysis, we find the scattering rates of 3ph, 4ph, 5ph, and 6ph processes approximately follows the $T^{1}$, $T^{2}$, $T^{3}$, and $T^{4}$, respectively.

However, temperature dependence alone cannot explain why 5ph and 6ph scatterings are substantially stronger in BaO than in MgO, particularly given that both materials adopt the same crystal structure and exhibit similar chemistries. The significantly enhanced 5ph and 6ph scatterings observed in BaO at room temperature can be quantified by computing the ratio of mode-averaged 5ph and 6ph scattering rates between MgO and BaO, respectively, as shown in Fig.~\ref{fig:ana}(b). The results reveal that 5ph scattering in BaO is more than one order of magnitude stronger than in MgO, while 6ph scattering is more than two orders of magnitude stronger than in MgO.

To explore the underlying physical origin, we analyze and derive approximated forms of Eq.~(\ref{eq:g5ph}) and (\ref{eq:g6ph}), i.e., $\gamma_{(n)\rm ph} \propto ||\Phi^{(n)}||^2/||\Phi^{(2)}||^{(n-1/2)}$, using only various orders of IFCs ($\Phi^{(n)}$) and simplifying the formula by noting (1) $|V^{(n)}|^2 \propto ||\Phi^{(n)}||^2/\omega^n$, (2) $n \propto k_{\rm B}T/(\hbar\omega)$, and (3) $\omega \propto ||\Phi^{(2)}||^{1/2}$. The estimated ratio of 5ph and 6ph scattering between MgO and BaO is shown in Fig.~\ref{fig:ana}(c). We find that our estimation semi-quantitatively agrees with Fig.~\ref{fig:ana}(b), revealing the dominant role of IFCs in shaping the relative strength of 5ph and 6ph scatterings. We note that our goal here is not to achieve quantitative accuracy, as we have adopted several drastic approximations, such as neglecting the scattering phase space arising from phonon band topology and the detailed spatial distribution of IFCs. These results demonstrate that the simple ratio of IFCs provides a promising framework for efficiently assessing the significance of 5ph and 6ph scatterings, for example, across all rocksalt and zincblende compounds~\cite{rczbprx}.

To illustrate the critical role of harmonic IFCs in determining the scattering rates, we also plot the scattering phase space by assuming $|V|=1$ in Eq.~(\ref{eq:g5ph}) and ~(\ref{eq:g6ph}) at 300~K, as shown in Fig.~\ref{fig:ana}(d). We find the scattering phase spaces associated with 5ph and 6ph processes in BaO are exceptionally large compared to Si and MgO. Additionally, 5ph and 6ph phase spaces are comparable in BaO, while they differ by orders of magnitudes in Si and MgO. Overall, the scattering phase space increases with decreasing Debye temperature, implying that stronger higher-order scattering is more likely to occur in softened lattice with weak harmonic interactions.

\begin{figure}[htp]
	\includegraphics[width = 0.9\linewidth]{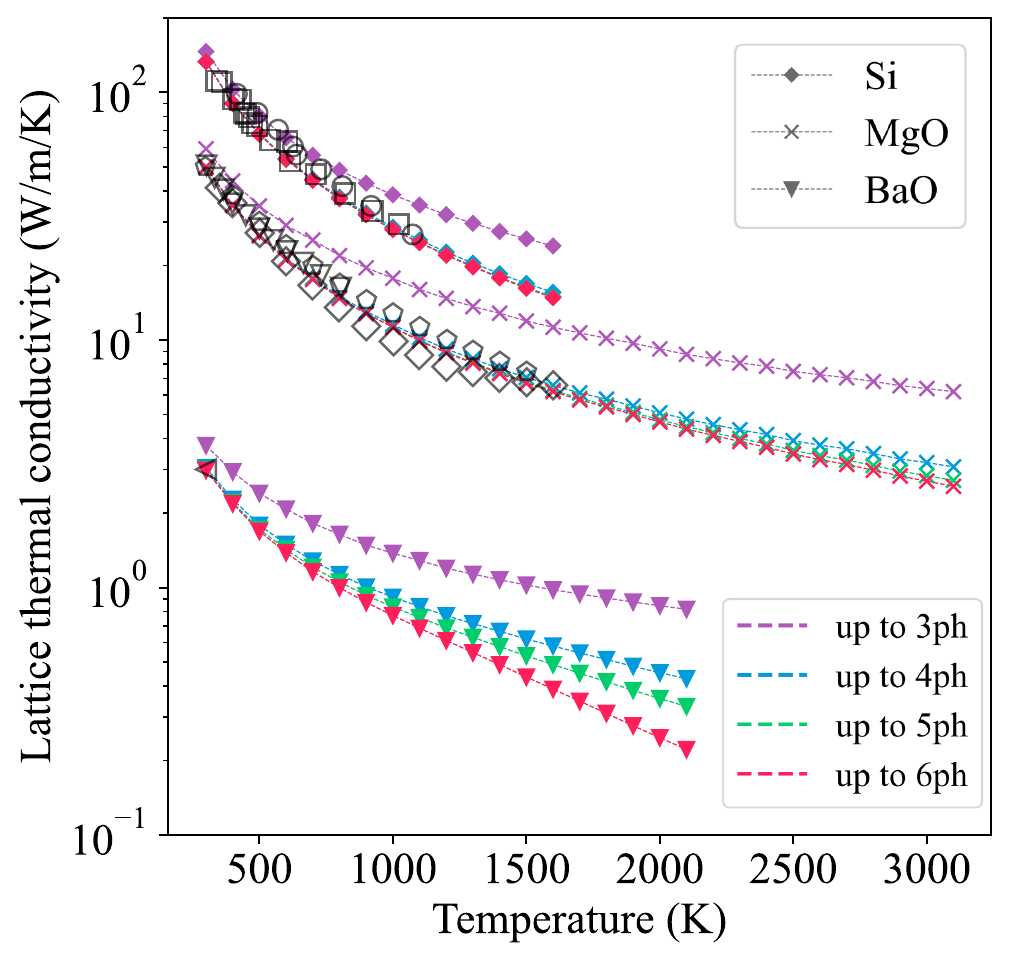}
	\caption{
	Temperature-dependent lattice thermal conductivities calculated with inclusion of up to three-phonon (3ph), four-phonon (4ph), five-phonon (5ph), and six-phonon (6ph) scattering processes for Si, MgO, and BaO, respectively. Experimental measurements are shown as empty symbols (squares from Ref.~[\onlinecite{abeles1962thermal}]; circles from Ref.~[\onlinecite{glassbrenner1964thermal}]; diamonds from Ref.~[\onlinecite{touloukion1970thermophysical}]; triangles from Ref.~[\onlinecite{cahill1998thermal}]; pentagons from Ref.~[\onlinecite{hofmeister2014thermal}]).
	}
	\label{fig:kappa}
\end{figure}

These scattering rates can be better understood through their impact on thermal transport properties, therefore we compute $\kappa_{\rm L}$ for all compounds from 300 K to temperatures near their melting points. Specifically, we adopted the unified theory of thermal transport in crystals and glasses~\cite{Simoncelli2019}, incorporating both the Peierls contribution~\cite{peierls1997kinetic} and the wavelike tunneling contribution~\cite{simoncelli2022wigner} to the heat current operators (see Supplementary Materials~\cite{SM} for calculation details), with these results shown in Fig.~\ref{fig:kappa}. The impact of 5ph and 6ph scattering is negligible in Si and MgO below 1500~K (4.4\% and 2.1\% reduction in $\kappa_{\rm L}$ for Si and MgO, respectively, at 1500~K), consistent with results in the literature that 3ph and 4ph processes capture the major physics in such temperature ranges~\cite{Tianli2017,kwon2020dominant,han2023predictions}. With increased temperature approaching the melting temperature, MgO shows further reduction due to 5ph and 6ph scatterings, with a percentage reduction of about 17\% at 3100~K on top of the combined 3ph and 4ph scatterings. In contrast, the reduction in $\kappa_{\rm L}$ for BaO at high temperatures is significant: 5ph and 6ph scatterings are able to reduce $\kappa_{\rm L}$ from 0.43~W/(m$\cdot$K) to 0.22~W/(m$\cdot$K) at 2100 K, demonstrating their dominating scattering effects. 

We emphasize that our main purpose in computing $\kappa_{\rm L}$ is to intuitively reflect the relative strength of multi-phonon scattering processes considered in our work, as successfully demonstrated above. We acknowledge the potential limitations for precisely determining $\kappa_{\rm L}$ using the present framework. These limitations include the absence of phonon frequency shifts (anharmonic phonon renormalization)~\cite{Errea2014, TDEP, Tadano2015, pbte2018}, the potential breakdown of the quasiparticle approximation~\cite{coiana2024breakdown,dangic2025lattice}, and the neglect of thermal expansion effects and temperature-dependent anharmonic force constants~\cite{li2023first,pbte2018,ravichandran2018unified}, all of which warrant rigorous investigation in future studies. Meanwhile, applying our formalism at very high temperatures may invoke non-analytic potentials in real materials, potentially challenging the current framework based on analytic Taylor expansions. This could also provide an opportunity to explore the limits and applicability of existing methodologies.

To summarize, we have achieved two main objectives in this study by first developing a theoretical formalism for describing five- and six-phonon scatterings and second studying multi-phonon scatterings in representative materials from first-principles.  
Looking ahead, future work should involve more systematic benchmarking against molecular dynamics simulations and experimental measurements, such as spectral functions and dynamical structure factors. Additionally, the recently developed mode coupling theory~\cite{castellano2023mode,castellano2025mode} could be explored to compute scattering rates using renormalized interatomic force constants. Furthermore, future theoretical developments should systematically evaluate other multi-vertex diagrams beyond the double-vertex classification to establish a complete perturbative framework for anharmonic phonon interactions. These combined efforts would contribute to a more comprehensive understanding of higher-order phonon interactions and their roles in thermal transport across diverse materials and extreme conditions.


\begin{acknowledgments}
\textbf{Acknowledgments:} 
Y. X. acknowledges 1) the support from the US National Science Foundation through awards DMR-2317008 and CBET-2445361, 2) the support from the Faculty Development Program at Portland State University, and 3) the computing resources provided by Bridges2 at Pittsburgh Supercomputing Center (PSC) through allocations mat220006p and mat220008p from the Advanced Cyber-infrastructure Coordination Ecosystem: Services \& Support (ACCESS) program, which is supported by National Science Foundation grants 2138259, 2138286, 2138307, 2137603, and 2138296. Y. X. is grateful to Z.J. W. for her encouragement and support during the preparation of this manuscript.
\end{acknowledgments}
\bibliography{CuSbS}

\end{document}